\documentclass{ws-procs975x65}

\begin{document}

\title{STRONGLY MAGNETIZED ELECTRON DEGENERATE GAS: HIGHLY SUPER-CHANDRASEKHAR WHITE DWARFS}

\author{UPASANA DAS$^*$ and BANIBRATA MUKHOPADHYAY$^\dag$}

\address{Department of Physics, Indian Institute of Science,\\
Bangalore 560012, India\\
E-mail: $^*$upasana@physics.iisc.ernet.in , $^\dag$bm@physics.iisc.ernet.in}

\begin{abstract}
We consider a relativistic, degenerate, electron gas under the influence of a 
strong magnetic field, which describes magnetized white dwarfs. 
Landau quantization changes the density of states available to the electrons, 
thus modifying the underlying equation of state. We obtain the mass-radius relations for 
such white dwarfs and show that it is possible
to have magnetized white dwarfs with a mass significantly greater than
the Chandrasekhar limit in the range $2.3 - 2.6 M_{\odot}$. Recent
observations of peculiar type~Ia supernovae - SN 2006gz, SN 2007if, SN
2009dc, SN 2003fg - seem to suggest super-Chandrasekhar-mass white dwarfs
with masses up to $2.4 - 2.8 M_{\odot}$, as their most likely progenitors and
interestingly our results lie within the observational limits.
\end{abstract}


\bodymatter

\section{Introduction}\label{aba:sec1}

Chandrasekhar first obtained the maximum possible mass for a stable white dwarf (WD) to be $\sim 1.44M_{\odot}$ \cite{chandra}, $M_\odot$ being the mass of Sun. Later, several magnetized WDs have been discovered with surface fields $10^{5}-10^{9}$ G \cite{vanlan}. It is likely that stronger fields ($\sim 10^{12}$ G) exist at their centers, which, however, cannot be probed directly. High magnetic field strength modifies the equation of state of the degenerate matter by causing Landau quantization of the electrons. In the present work, we study the effect of magnetic field strengths $\gtrsim 10^{15}$ G on the electron degenerate matter and hence obtain the modified mass-radius relation of the strongly magnetized WD. Interestingly, we find that it is possible to have a super-Chandrasekhar WD having mass $\sim 2.3 - 2.6 M_{\odot}$, provided it has an appropriate magnetic field strength and central density. Incidentally, there have been recent observations of peculiar type~1a supernovae - SN2006gz, SN2007if, SN2009dc - which invoke super-Chandrasekhar WDs of mass up to $2.4-2.8M_{\odot}$ as their progenitors \cite{scalzo}. These observations remarkably concur with the 
super-Chandrasekhar WDs obtained in this work.

\section{Theory}

The energy states of a free electron in a uniform magnetic field $B$ are quantized into Landau orbitals, which defines the motion of the electron in a plane perpendicular to $B$. The relevant equations (see Ref. 4) are obtained by solving the Dirac equation in a strong magnetic field, such that $B \ge B_{c} = 4.414 \times 10^{13}$ G. The maximum number of Landau levels occupied by a gas of cold electrons in a magnetic field is given by $\nu_{m} = \frac{(E_{\rm Fmax}/m_e c^2)^2-1}{2B_D}$, where $m_{e}$ is the rest mass of the electron, $c$ the speed of light, $E_{\rm Fmax}$ the maximum Fermi energy of the system and $B_{D}=B/B_c$. Thus a high $B_D$ corresponds to a low $\nu_{m}$, i.e., a smaller number of Landau levels are occupied, which is when the magnetic field plays an important role in modifying the equation of state (EoS) of the relativistic electron degenerate gas.


Next, in order to obtain the new mass-radius relation for a strongly magnetized WD, we combine the modified EoS with the condition of magnetostatic equilibrium. If the WD is approximated to be spherical, then its mass $M$ is obtained from:
\begin{equation}
\frac{1}{\rho}\frac{d}{dr}\left(P+\frac{B^2}{8\pi}\right)=-\frac{GM}{r^2}+\left[\frac{\vec{B}\cdot\nabla\vec{B}}{4\pi\rho}\right]_r ~{\rm and}~~ \frac{dM}{dr} = 4\pi r^{2}\rho,
\label{mass}
\end{equation}
where $r$ is the radial distance from the center of the WD, $P$ and $\rho$ are the pressure and density of the electron degenerate matter respectively and $G$ is Newton's gravitation constant.
However, if the magnetic field is
very strong, then the pressure becomes anisotropic \cite{sinha} and also a magnetic tension develops in the combined fluid-magnetic 
medium \cite{bocquet}, which flattens the WD
along the direction of $B$.
We estimate the effect of deviation from spherical symmetry by assuming a cylindrical geometry, such that $M$ is now obtained from:
\begin{equation} 
\frac{1}{\rho}\frac{d}{dr_{eq}}\left(P+\frac{B^2}{8\pi}\right) = -\frac{GM}{r^{2}}\left(\frac{r_{eq}}{r}\right)+ \left[\frac{\vec{B}\cdot\nabla'\vec{B}}{4\pi\rho}\right]_{r_{eq}} {\rm and}~~ \frac{dM}{dr_{eq}} = 2\pi r_{eq} h \rho,
\label{Meq}
\end{equation}
where $r_{eq}$ is the equatorial distance from the center of the WD and  $h$ denotes an average height of the cylinder, which also quantifies the degree of flattening \cite{dasmu} (note $\nabla$ and $\nabla'$ differ according to the chosen co-ordinate systems). At high $B$, the WD will be more flattened and $h$ will be small, whereas at low $B$ it will be less flattened and hence $h$ will be large. Note that in this work our interest is to see the effect of the magnetic field of the central region of the WD, where the field is supposedly (almost) constant. Hence, the gradients of $B$ are negligible in Eqs. (1) and (2). See Ref. 5 for details. Moreover, any currents generated near the surface, where $B$ varies, would not affect our main conclusion.

\section{Results}

\def\figsubcap#1{\par\noindent\centering\footnotesize(#1)}
\begin{figure}[h]
\begin{center}
\parbox{2.1in}{\epsfig{figure=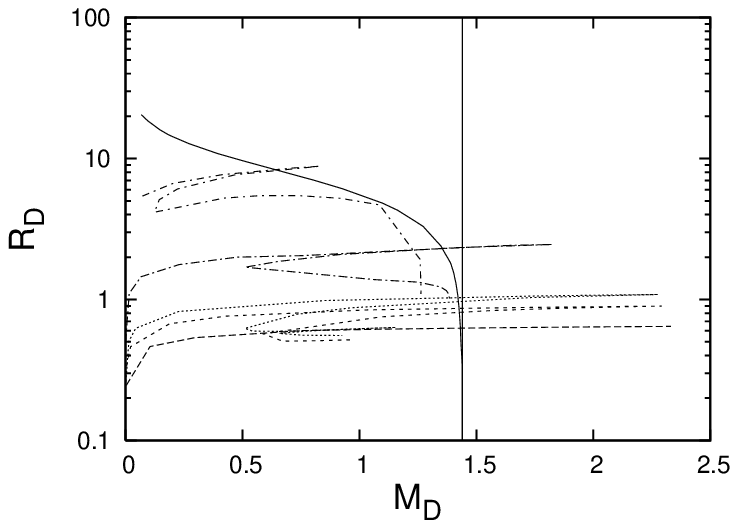,width=2in}
\figsubcap{a}}
 \hspace*{4pt}
 \parbox{2.1in}{\epsfig{figure=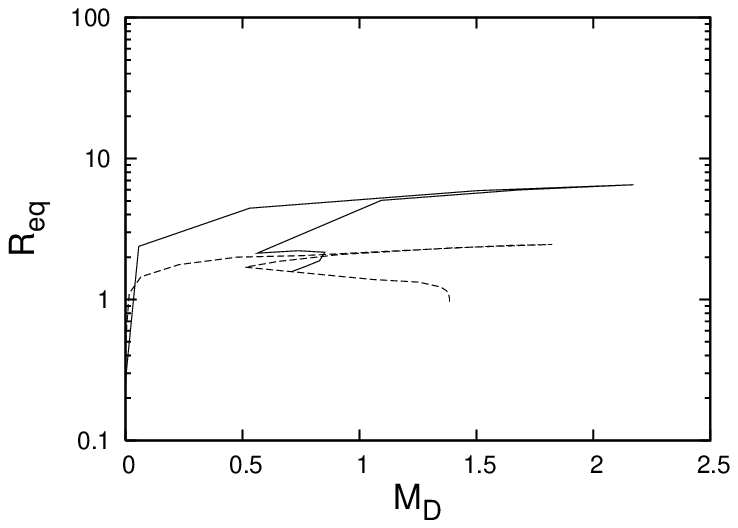,width=2in}
 \figsubcap{b}}
 \caption{(a) Spherical case - the solid line represents Chandrasekhar's mass-radius relation and the vertical line marks the 1.44$M_{\odot}$ limit. From top to bottom the other lines represent spherical WDs corresponding to $\nu_{m} = 500\,, 20\,, 3\,, 2$ and $1$ respectively. (b) Flattened case - the solid line represents the mass-radius relation for flattened WDs corresponding to $\nu_{m} =20$, while the dashed line represents the same for the corresponding spherical WD. ${\rm M_D}$ is the mass of the WD in units of $M_{\odot}$; ${\rm R_D}$ and ${\rm R_{eq}}$ are in units of $10^{8}$ cm. $E_{\rm Fmax} = 20\,m_{e}c^{2}$.}
\label{fig1.2}
\end{center}
\end{figure}

Figure 1(a) shows the mass-radius relations for spherical WDs corresponding to different magnetic field strengths, along with Chandrasekhar's mass-radius relation. We observe that as $B$ decreases, or equivalently as $\nu_m$ increases, the mass-radius relations approach Chandrasekhar's relation. Also, as $B$ increases, the WDs become more compact in size and have larger masses. We note that the most massive WD, corresponding to a 1-level system (with $B = 8.8\times10^{15}$ G and $\nu_m=1$), has $M \sim$ 2.3 $M_{\odot}$. 

Figure 1(b) shows the mass-radius relation for the flattened WDs corresponding to a 20-level system, along with the one for the corresponding spherical case. We observe that the flattened mass-radius relation exhibits a higher WD mass, with a maximum mass $\sim 2.2 M_\odot$, compared to the spherical one, which has a maximum mass $\sim 1.8 M_\odot$. Thus, even for a much lower magnetic field strength (with $B = 4.4\times10^{14}$ G for $\nu_{m}=20$), WDs having $M \gtrsim 2 M_\odot$ are obtained if one includes the appropriate flattening effect. Note that the radius for the flattened case is the equatorial radius (${\rm R_{eq}}$), which is automatically larger than the spherical radius (${\rm R_D}$) as a consequence of the flattening effect.


\section{Conclusion}

We find that the effect of Landau quantization due to a strong magnetic field on the underlying electron degenerate matter gives rise to significantly super-Chandrasekhar WDs \cite{dasmu}. The maximum mass obtained for magnetized WDs is about $\sim 2.3 M_{\odot}$, corresponding to a central magnetic field $8.8\times 10^{15}$ G. Increasing the magnetic field leads to masses up to $2.6M_\odot$. However, the magnetic tension due to strong magnetic field causes a deformation in the WDs which adopt a flattened shape. Incorporating this flattening effect leads to more massive WDs (compared to the corresponding spherical ones) even at relatively lower magnetic field strengths - such relatively weakly magnetized white dwarfs are more probable in nature. \\

This work was partly supported by the ISRO grant ISRO/RES/2/367/10-11. U.D. thanks CSIR, India for financial support.


\end{document}